\documentclass[twocolumn]{emulateapj09}

\usepackage{graphicx}
\usepackage{epsfig}
\usepackage{amsmath}

\shorttitle{Stars around Sgr~A*}
\shortauthors{Psaltis, Li, and Loeb}

\newcommand{\sgra}{Sgr~A$^*$}

\begin{document}


\title{Deviation of Stellar Orbits from Test Particle Trajectories\\
  Around Sgr~A* Due to Tides and Winds}

\author{Dimitrios Psaltis\altaffilmark{1,2}, Gongjie
  Li\altaffilmark{2}, and Abraham Loeb\altaffilmark{2}}

\altaffiltext{1}{Astronomy Department,
University of Arizona,
933 N.\ Cherry Ave.,
Tucson, AZ 85721, USA}

\altaffiltext{2}{Institute for Theory \& Computation, Harvard-Smithsonian
  CfA, 60 Garden Street, Cambridge, MA, USA}

\email{dpsaltis@email.arizona.edu; gli@cfa.harvard.edu;
  aloeb@cfa.harvard.edu}

\begin{abstract}
Monitoring the orbits of stars around \sgra\ offers the possibility of
detecting the precession of their orbital planes due to frame
dragging, of measuring the spin and quadrupole moment of the black
hole, and of testing the no-hair theorem. Here we investigate whether
the deviations of stellar orbits from test-particle trajectories due
to wind mass loss and tidal dissipation of the orbital energy
compromise such measurements. We find that the effects of stellar winds
are, in general, negligible. On the other hand, for the most eccentric
orbits ($e>0.96$) for which an optical interferometer, such as
GRAVITY, will detect orbital plane precession due to frame dragging,
the tidal dissipation of orbital energy occurs at timescales
comparable to the timescale of precession due to the quadrupole moment
of the black hole. As a result, this non-conservative effect is a
potential source of systematic uncertainty in testing the
no-hair theorem with stellar orbits.
\end{abstract}


\section{INTRODUCTION}

Stars in orbit around the black hole in the center of the Milky Way,
hereafter \sgra, have been tracked for more than a decade, providing a
measure of the black hole mass (Genzel et al.\ 2010; Ghez et
al.\ 2012). The constraints have been steadily improving with the
first measurement of a fully closed orbit for the star S2 (see, e.g.,
Ghez et al.\ 2008; Gillessen et al.\ 2009) as well as with the
discovery of additional stars (S0-16, S0-102 and S0-104) in
orbits that probe the black-hole spacetime within a few thousand
gravitational radii (Meyer et al.\ 2012).

Precise astrometric observations of stars in close orbits around
\sgra\ may lead to the detection of orbital precession due to general
relativistic frame dragging, measuring the spin of the black hole,
and testing the no-hair theorem (Will 2008).  Such measurements
will be complementary to those that will be achieved with the Event
Horizon Telescope (Fish \& Doeleman 2009; Johannsen \& Psaltis 2010)
as well as to timing observations of pulsars in orbit around the black
hole (Pfahl \& Loeb 2004; Liu et al.\ 2012).

Future instruments, such as GRAVITY, an adaptive-optics assisted
interferometer on the VLT (Eisenhauer et al.\ 2011), will track
stellar orbits with a single pointing astrometric accuracy of $\simeq
10-200~\mu$arcsec, for stars as faint as $m_{\rm K}=16.3-18.8$ in a
crowded field (Stone et al. 2012). At this resolution, the biggest
challenge in measuring the fundamental properties of \sgra\ with
stellar orbits will be ensuring that a particular measurement is not
affected adversely by astrophysical complications.

A number of studies have explored the effects of non-gravitational
forces exerted on the orbiting stars by other objects in the same
environment.  Merritt et al\ (2010) and Sadeghian \& Will (2011)
investigated the perturbative effects of the stellar cluster on the
orbits of individual stars and found that they are negligible compared
to the general relativistic effects inside $\sim$1~mpc$\simeq 5\times
10^3$ gravitational radii. Psaltis (2012) studied the interaction of
the orbiting stars with the ambient gas and showed that hydrodynamic
drag and star-wake interactions are negligible inside $\sim
10^5$~gravitational radii. 

In this paper, we study the deviations of the stellar orbits from
test-particle trajectories that are introduced by the fact that stars
are not point particles but {\em (i)\/} may lose mass in strong winds
and {\em (ii)} may be tidally deformed. We calculate the range of
orbital parameters for which orbital perturbations due to the stellar
winds and tides do not preclude the measurement of the black-hole spin
and quadrupole moment and, therefore, testing of the no-hair theorem.

\section{Characteristic Timescales}

We start by comparing the characteristic timescales for orbital
precession due to general relativistic effects to those of orbital
perturbations due to stellar winds and to tidal forces.  Hereafter, we
set the mass of the black hole to $4\times 10^6 M_\odot$ and its
distance to 8.4~kpc. We also denote by $M_{\rm BH}$ the mass of the
black hole, by $M_{\rm S}$ the mass of the star, and by $a$ and $e$
the semi-major axis and eccentricity of the stellar orbit. With these
definitions, the Newtonian period of a stellar orbit is
\begin{eqnarray}
P&=&2\pi\left(\frac{a^3}{GM_{\rm BH}}\right)^{1/2}\nonumber\\
&=&123.8\left(\frac{M_{\rm BH}}{4\times 10^6~M_\odot}\right)
\left(\frac{ac^2}{GM_{\rm BH}}\right)^{3/2}~\mbox{s}\;.
\end{eqnarray}

\subsection{Dynamical Timescales}

General relativistic corrections to Newtonian gravity affect the
orbits of stars around \sgra\ in, at least, three ways. 

First, eccentric orbits precess on the orbital plane (periapsis
precession). The characteristic timescale for this precession is
(Merritt et al.\ 2010)
\begin{eqnarray}
t_{\rm S}&=&\frac{P}{6}\frac{c^2 a}{GM_{\rm BH}}\left(1-e^2\right)\nonumber\\
&=&20.63\left(\frac{M_{\rm BH}}{4\times 10^6~M_\odot}\right)
\left(\frac{ac^2}{GM_{\rm BH}}\right)^{5/2}\left(1-e^2\right)~\mbox{s}\;.
\end{eqnarray}

Second, orbits with angular momenta that are not parallel to the 
spin angular momentum of the black hole precess because of frame
dragging. The characteristic timescale for this precession is
(Merritt et al.\ 2010)
\begin{eqnarray}
t_{\rm J}&=&\frac{P}{4\chi}
\left[\frac{c^2a\left(1-e^2\right)}{G M_{\rm BH}}\right]^{3/2}\nonumber\\
&=&
30.95\chi^{-1}\left(\frac{M_{\rm BH}}{4\times 10^6~M_\odot}\right)
\left(\frac{ac^2}{GM_{\rm BH}}\right)^{3}\left(1-e^2\right)^{3/2}~\mbox{s}
\;,\nonumber\\
\end{eqnarray}
where $\chi$ is the spin of the black hole.

Finally, tilted orbits also precess because of the quadrupole moment 
of the spacetime. The characteristic timescale for this precession is
(Merritt et al.\ 2010)
\begin{eqnarray}
t_{\rm Q}&=&\frac{P}{3\vert q\vert}
\left[\frac{c^2a\left(1-e^2\right)}{G M_{\rm BH}}\right]^2\nonumber\\
&=&
41.26\vert q\vert^{-1}\left(\frac{M_{\rm BH}}{4\times 10^6~M_\odot}\right)
\left(\frac{ac^2}{GM_{\rm BH}}\right)^{7/2}\left(1-e^2\right)^{2}~\mbox{s}
\;,\nonumber\\
\end{eqnarray}
where $q$ is the quadrupole moment of the black-hole spacetime. If the
spacetime of the black hole satisfies the no hair theorem, then
$q=-\chi^2$.

The three timescales for a spinning Kerr black hole ($\chi=0.3$,
$q=-\chi^2$) and for orbits with two different eccentricities are
shown in Figure~1 as a function of the orbital semi-major axis.

\subsection{Wind mass loss}

The angular momentum of a star in orbit around the black hole is
\begin{equation}
J=M_{\rm S}\left(GM_{\rm BH}a\right)^{1/2}
\end{equation}
(assuming here for simplicity a circular orbit). If the star is
losing mass in a wind at a rate $\dot{M}_{\rm w}$, then its orbit
will evolve according to
\begin{equation}
\frac{\dot{a}}{a}=
2\frac{\dot{J}}{J}-2\frac{\dot{M}_{\rm w}}{M_{\rm S}}\;.
\end{equation}
Assuming that the wind is carrying a fraction $\eta$ of the orbital
angular momentum, i.e.,
\begin{equation}
\dot{J}_{\rm w}=\eta \dot{M}_{\rm w}\left(GM_{\rm BH}a\right)^{1/2}
\end{equation}
then the rate of change of the orbital separation becomes
\begin{equation}
\frac{\dot{a}}{a}=2(1-\eta)\frac{\dot{M}_{\rm w}}{M_{\rm S}}\;.
\end{equation}
In other words, the timescale for orbital evolution due to the presence
of the wind is
\begin{equation}
\tau_{\rm w}\equiv\left\vert\frac{a}{\dot{a}}\right\vert=
\left[\frac{1}{2(1-\eta)}\right]
\frac{M_{\rm S}}{\dot{M}_{\rm w}}\;,
\end{equation}
or
\begin{equation}
\tau_{\rm w,-7}=1.6\times 10^{15}\left(\frac{1}{1-\eta}\right)
\left(\frac{M_{\rm S}}{10 M_\odot}\right)
\left(\frac{\dot{M}_{\rm w}}{10^{-7} M_\odot~\mbox{yr}^{-1}}\right)^{-1}~\mbox{s}\;,
\nonumber\\
\end{equation}
where we have used the subscript ``-7'' to denote the exponent in the
wind mass loss rate.

This characteristic timescale is compared to the dynamical timescales
in Figure~1, for a $10~M_\odot$ star and for a wind mass-loss rate of
$10^{-7} M_\odot$~yr$^{-1}$, which is consistent with current
observations of the star S2 in orbit around \sgra\ (Martins et
al.\ 2008). The effect of wind mass loss becomes negligible with
respect to the frame-dragging induced precession of the orbital planes
for orbits within $\sim 30,000$ gravitational radii.  On the other
hand, they become negligible with respect to the quadrupole induced
precession of the orbital planes for orbits within $\sim 4,000$
gravitational radii.

\begin{figure}[t]
\psfig{file=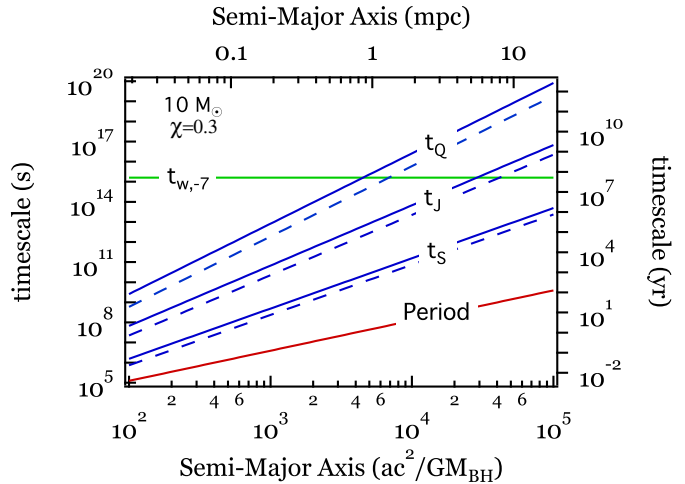,width=3.5in}
\caption{Different timescales that are relevant to the evolution of
  orbits of stars in the vicinity of \sgra, as a function of their
  semi-major axes. The red line shows the periods of the orbits. The
  blue lines show the timescales for the precession of the periapsis
  ($t_{\rm S}$), for the precession of the orbital plane due to frame
  dragging ($t_{\rm J}$), and for the precession of the orbital frame
  due to the quadrupole moment of the spacetime ($t_{\rm Q}$); the
  black-hole spin is taken to be $\chi=0.3$ and solid and dashed lines
  correspond to eccentricities of 0.5 and 0.8, respectively. The green
  line ($t_{\rm w,-7}$) shows the characteristic timescale for orbital
  evolution of a $10 M_\odot$ star due to the presence of a stellar
  wind at a mass loss rate of $10^{-7} M_\odot$~yr$^{-1}$, for
  $\eta$ set equal to zero.}
\label{fig:winds}
\end{figure}

\subsection{Tidal Dissipation of Orbital Energy}

\begin{figure}[t]
\psfig{file=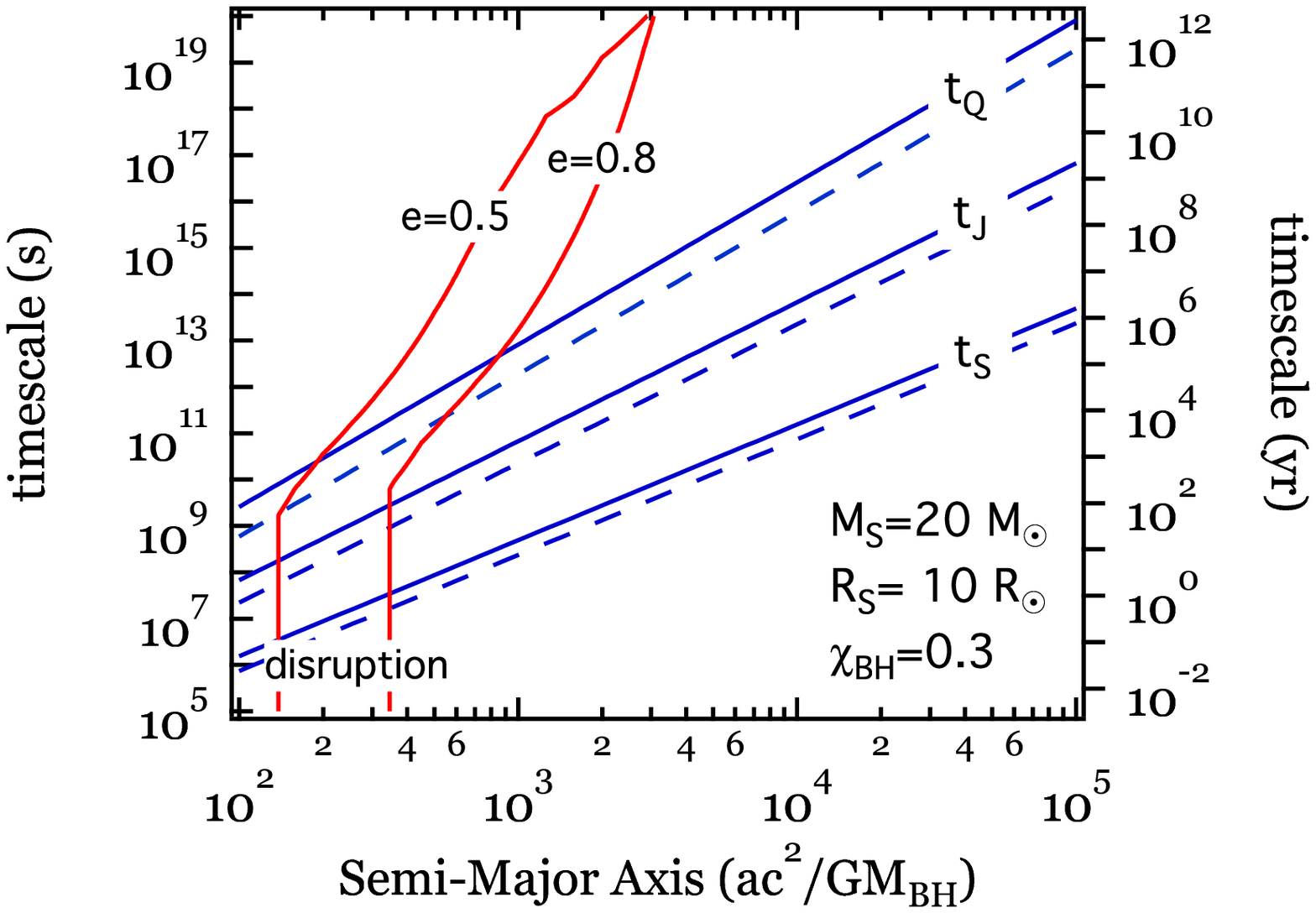,width=3.5in}
\psfig{file=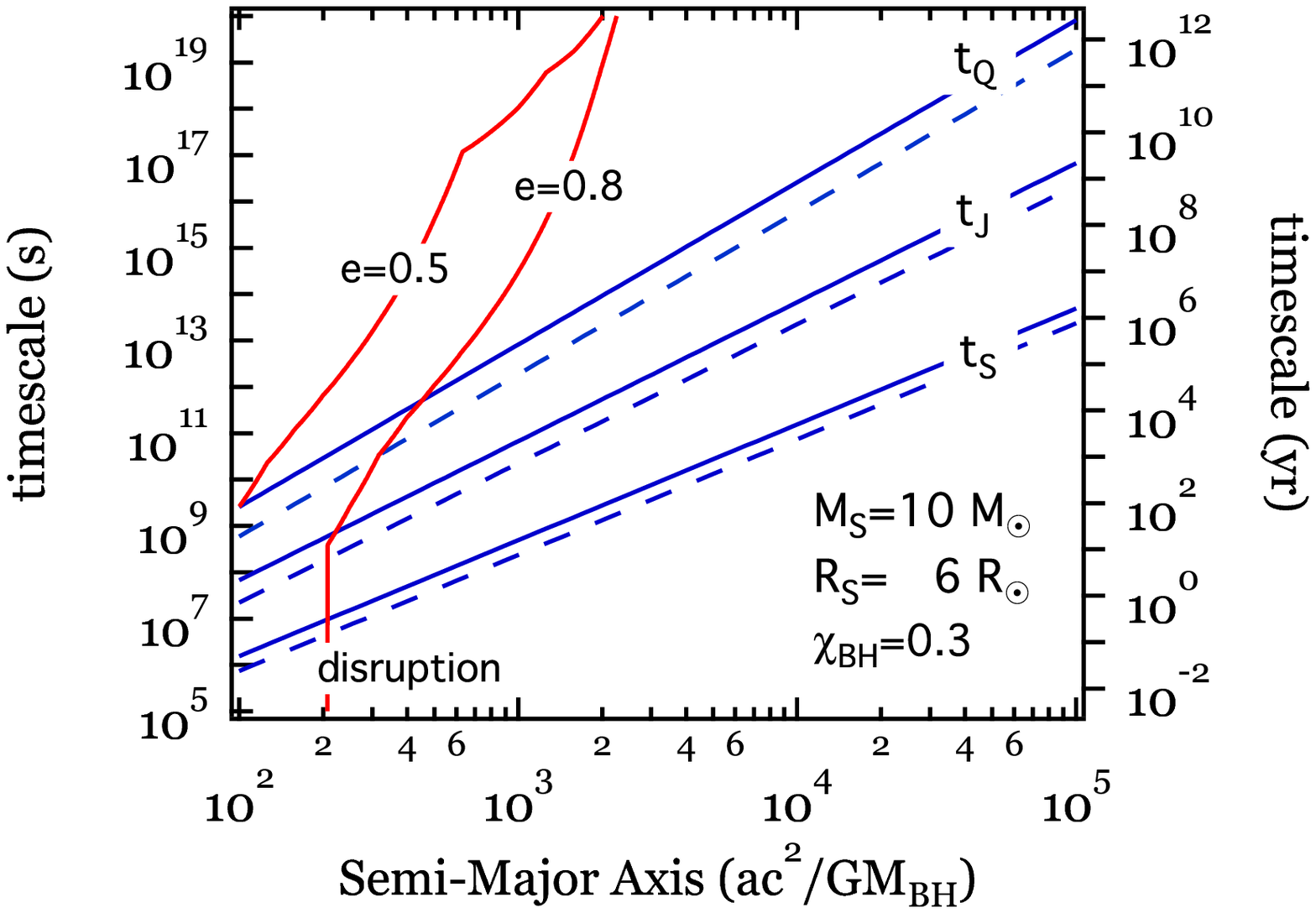,width=3.5in}
\caption{The blue lines show the dynamical timescales, as in
  Figure~1. The red lines show the characteristic timescale for
  orbital evolution due to the tidal dissipation of the orbital
  energy, for two different values of the eccentricity. The vertical
  segments of the red lines indicate the semi-major axes at which the
  stars are tidally disrupted at periastron. The two panels correspond
  to a 20~$M_\odot$ and a 10~$M_\odot$ star. In both cases, the
  black-hole spin is taken to be equal to $\chi=0.3$.}
\label{fig:tides}
\end{figure}

The tidal deformations excited at each periastron passage transfer
some of the orbital energy into modes within the volume of the star
(see Alexander 2006 for a review of stellar processes around
  \sgra). Since the orbital energy loss is proportional to the number
of passages (Li \& Loeb 2012), we can use the approach of Press \&
Teukolsky (1977) to estimate the rate of dissipation of orbital energy
as
\begin{equation}
\frac{\Delta E}{\Delta t}\simeq
\left(\frac{GM_{\rm S}^2}{PR_{\rm S}}\right)
\left(\frac{M_{\rm BH}}{M_{\rm S}}\right)^2
\sum_{l=2,3,...} \left(\frac{R_{\rm S}}{R_{\rm p}}\right)^{2l+2}T_l(\eta)\;.
\end{equation}
Here $R_{\rm p}=a(1-e)$ is the periastron distance, $R_{\rm S}$ is the
radius of the star, and $T_l(\eta)$ are appropriate dimensionless
functions of the quantity 
\begin{equation}
\eta\equiv \left(\frac{M_{\rm S}}{M_{\rm S}+M_{\rm BH}}\right)^{1/2}
\left(\frac{R_{\rm p}}{R_{\rm S}}\right)^{3/2}
\end{equation}
that describe the excitation of modes with different spherical
harmonic index $l$. 

In detail,
\begin{equation}
T_l(\eta) = 2 \pi^2 \sum_{n, m} |Q_{nl}|^2 |K_{nlm}|^2\;,
\label{e:E0n} 
\end{equation}
where $n$ is the mode order and $m$ is the other spherical harmonic
index. The excited modes have $l>1$ and $-l<m<l$.  The
coefficient $K_{nlm}$ represents the coupling to the orbit,
\begin{equation}
K_{nlm} = \frac{W_{lm}}{2\pi} \int _{-\infty}^{\infty} dt \left[
\frac{R_p}{r(t)}\right]^{l+1} \rm{exp} \{i [ \omega_n t + m {\Phi
(t)} ] \} ,
\end{equation}
where $r(t)$ is the instantaneous distance between the star and \sgra,
$\omega_n$ is the mode frequency, $\Phi (t)$ is the true anomaly, and
\begin{eqnarray}
W_{lm} &=& (-1)^{(l+m)/2}\left[\frac{4 \pi}{(2 l+1)}(l-m)!(l+m)!\right] ^{1/2}
\nonumber\\
&&\qquad
\left[2^l\frac{(l-m)}{2}!\frac{(l+m)}{2}!\right]^{-1}\;.
\end{eqnarray}  

The tidal overlap integral $Q_{nl}$ represents the coupling of the
tidal potential to a given mode, i.e.,
\begin{equation}
Q_{nl} = \int_0^1 R^2 dR \rho(R) l R^{l-1} [\xi_{nl}^{\cal R} +
  (l+1)\xi_{nl}^{\cal S}]\;,
\end{equation}
where $\rho(R)$ is the stellar density profile as a function of radius
$R$ and $\xi (R) = [\xi_{nl}^{\cal R} (R) \hat{e}_R + \xi_{nl}^{\cal
S} (R) R \nabla] Y_{lm} (\theta, \phi) $ is the mode eigenfunction,
with $\xi_{nl}^{\cal R}$ and $\xi_{nl}^{\cal S}$ being its radial and
poloidal components, respectively. We obtain the appropriate stellar
density profile from the MESA code (Paxton et al.\ 2011) and compute
the mode eigenfunctions with the ADIPLS code (Christensen-Dalsgaard
2008).

Because the energy gain in each passage depends on
$({R_S}/{R_p})^{2l+2}$ and the values of $Q_{nl}$ and
$K_{nlm}$ are similar for modes with different values of $l$, the
quadrupole ($l=2$) modes gain the most energy during the tidal
excitation (the $l=0$ and $l=1$ modes are not excited). For this reason,
we focus, hereafter, on the $l=2$ modes.

The characteristic timescale for orbital evolution due to tidal dissipation
is
\begin{eqnarray}
t_{\rm d}&\equiv& \frac{E}{\Delta E/\Delta t}\nonumber\\
&=&\frac{\pi R_{\rm S}}{c}
\left(\frac{G M_{\rm BH}}{c^2 R_{\rm S}}\right)^6
\left(\frac{M_{\rm S}}{M_{\rm BH}}\right)
\left(\frac{ac^2}{GM_{\rm BH}}\right)^{13/2}(1-e)^6T_2^{-1}\nonumber\\
&=&1.37\times 10^{-4}\left(\frac{M_{\rm BH}}{4\times 10^6~M_\odot}\right)^{5}
\left(\frac{R_{\rm S}}{10~R_\odot}\right)^{-5}
\left(\frac{M_{\rm S}}{20~M_\odot}\right)\nonumber\\
&&\qquad
\left(\frac{ac^2}{GM_{\rm BH}}\right)^{13/2}\left(1-e\right)^{6}
T_2^{-1}(\eta)~\mbox{s}\;,
\end{eqnarray}
and is shown in Figure~\ref{fig:tides} for two main-sequence stars
with masses $10 M_\odot$ and $20 M_\odot$.

If the star at periastron reaches inside the tidal radius
\begin{equation}
R_{\rm t}=R_{\rm S}\left(\frac{M_{\rm BH}}{M_{\rm S}}\right)^{1/3}\;, 
\end{equation}
it gets disrupted. For simplicity, we ignore here the fact that, if
the periastron distance is smaller than $4-5$ times the tidal radius,
the repeated heating of the star at each passage will make it
vulnerable to tidal disruption (Li \& Loeb 2012). Requiring $R_{\rm
    p}\ge R_{\rm t}$ sets a lower limit on the semi-major axis of the
  stellar orbit, i.e.,
\begin{eqnarray}
\left(\frac{ac^2}{GM_{\rm BH}}\right)&\ge& \frac{68.9}{1-e}
\left(\frac{R_{\rm S}}{10~R_\odot}\right)\nonumber\\
&&\qquad
\left(\frac{M_{\rm BH}}{4\times 10^6~M_\odot}\right)^{-2/3}
\left(\frac{M_{\rm S}}{20~M_\odot}\right)^{-1/3}\;.
\end{eqnarray}
The tidal limit is shown as the vertical portion of the red lines in
Figure~\ref{fig:tides}. At orbital separations larger than this limit,
the tidal evolution of the stellar orbits is never fast enough to
compete with the precession of the orbital planes due to frame
dragging. On the other hand, the orbital plane precession due to the
quadrupole moment of the black hole for stars with semi-major axes a
few times larger than the tidal limit will be masked by the orbital
evolution due to tidal effects.

\begin{figure}[t]
\psfig{file=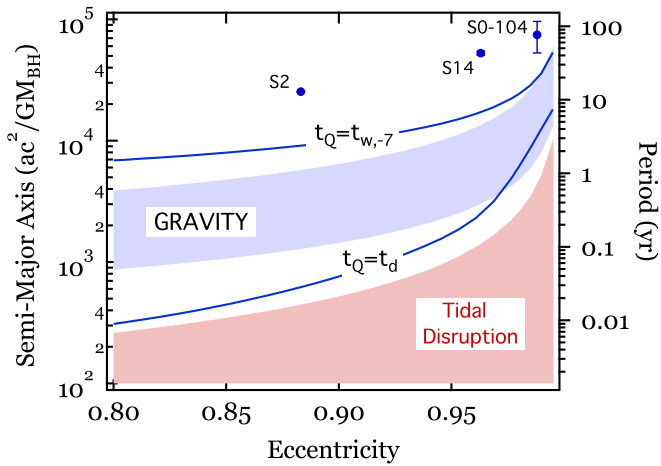,width=3.5in}
\caption{The two blue curves show the loci of orbital parameters for
  stars around \sgra\ at which the timescale of orbital-plane
  precession due to the quadrupole moment of the black hole ($t_{\rm
    Q}$) is equal to the orbital evolution timescale due to stellar
  winds ($t_{\rm w,-7}$) or due to tides ($t_{\rm d}$). In order
    for stars to follow nearly test-particle trajectories, their
    orbital parameters have to lie between the two curves.  The blue
  shaded area shows the range of orbital parameters for which frame
  dragging will be detectable with GRAVITY at a signal-to-noise ratio
  of 5, assuming a range of astrometric accuracies between
  $10-200~\mu$arcsec. The red shaded area show the range of orbital
  parameters that lead to the tidal disruption of the star at
  periapsis. All curves are for a black-hole spin of $\chi=0.3$ and a
  10~$M_\odot$ star. The three filled circles show the orbital
  parameters of the three stars nearest to \sgra\ that are presently
  known.}
\label{fig:orbits}
\end{figure}

\section{DISCUSSION}

We explored whether deviations of the orbits of star around \sgra\
from test particle trajectories due to stellar winds and tides may
compromise the measurements of relativistic
effects. Figure~\ref{fig:orbits} summarizes our results for an
illustrative case of a $10~M_\odot$ star and a black-hole spin of
$\chi=0.3$. The two blue curves in this figure show the combinations
of semi-major axes and orbital eccentricities for which the timescale
of orbital plane precession due to the quadrupole moment of the black
hole is equal to the orbital evolution timescale due to the wind-mass
loss ($t_{\rm w}=t_{\rm Q}$) and due to tides ($t_{\rm d}=t_{\rm
Q}$). In order for a stellar orbit not to be affected significantly by
either of the two effects, its parameters need to be in between the
two curves.

For comparison, we calculate the signal-to-noise ratio at which the
precession of the orbital plane of a star due to frame dragging will
be detected, in the near future, using the adaptive-optics assisted
interferometer GRAVITY. Following Weinberg et al.\ (2005), we write
the signal-to-noise ratio as
\begin{equation}
S=\frac{8\pi\chi}{a^{1/2}(1+e)^{1/2}(1-e)^{3/2}}
\left(\frac{G M_{\rm BH}}{Dc^2}\right)^{3/2}
\frac{N_{\rm orb} {\cos\psi}}{\delta \theta}\;,
\end{equation}
where $D$ is the distance to the black hole, $N_{\rm orb}$ is the
number of orbits monitored, $\cos\psi$ is the inclination of the
orbit, and $\delta\theta$ is the astrometric accuracy of each
measurement. Assuming that we monitor a particular orbit for a time
$\Delta T$, we can rewrite this expression as
\begin{eqnarray}
S&=&\frac{9\times 10^6\cos\psi}{(1+e)^{1/2}(1-e)^{3/2}}
\left(\frac{\chi}{0.3}\right)
\left(\frac{\Delta T}{10~\mbox{yr}}\right)
\left(\frac{D}{8.4~\mbox{kpc}}\right)^{-1}\nonumber\\
&&\qquad\qquad\left(\frac{\delta\theta}{10~\mu\mbox{arcsec}}\right)^{-1}
\left(\frac{ac^2}{GM_{\rm BH}}\right)^{-2}\;.
\end{eqnarray}
The astrometric accuracy of GRAVITY is expected to be $\sim
200~\mu$arcsec for a faint star of $m_{\rm K}\simeq 18.8$ and $\sim
10~\mu$arcsec for a brighter star of $m_{\rm K}=16.3$. Requiring a
signal-to-noise ratio of 5 for this range of astrometric accuracies
and for the typical parameters used in the above equation places an
upper limit on the semi-major axes of orbits as a function of their
eccentricity. This range of upper limits is shown as the blue-shaded
region in Figure~\ref{fig:orbits}. 

For all but the most eccentric orbits for which GRAVITY will be able
to detect orbital-plane precession due to frame dragging, both
effects of stellar winds and tides do not preclude by themselves the
measurement of the quadrupole moment of the black hole. On the other
hand, for highly eccentric orbits ($e>0.96$), the tidal dissipation of
orbital energy for massive stars occurs at similar timescales as the
orbital-plane precession due to the quadrupole moment of the black
hole. As a result, it needs to be taken into account as a possible
source of systematic uncertainties in measuring the quadrupole
moment of the black hole and in testing the no-hair theorem.

\acknowledgements

We thank S.\ Gillessen for his comments on the manuscript and F.\ \"Ozel
for many constructive disussions and comments. DP acknowledges the support
of the NSF CAREER award AST-0746549.  This work also was supported in
part by NSF grant AST-0907890 and NASA grants NNX08AL43G and
NNA09DB30A (for A.L.).

\end{document}